\documentclass{ws-ijmpa}

\usepackage{amssymb}
\usepackage{amsmath}
\usepackage{graphicx}
\usepackage{physics}
\usepackage{booktabs}
\usepackage{multirow}
\usepackage{makecell}
\usepackage{textcomp}
\usepackage{hyperref}
\usepackage{orcidlink}
\usepackage[table]{xcolor}
\usepackage{soul}
\newcommand{\rev}[1]{#1}
\hypersetup{hidelinks}

\begin{document}

\markboth{Shashank Verma, Manakkumar Parmar, and Ajay Kumar Rai}
{Mass Dependence in $B_c$ Decays}

%
\catchline{}{}{}{}{}
%

\title{Mass Dependence of Transition Form Factors and Decay Observables in 
$B_c \rightarrow \eta_c$
and 
$B_c \rightarrow J/\psi$
Decays}

\author{Shashank Verma \orcidlink{0009-0005-9312-1294}}
\address{Department of Physics, Sardar Vallabhbhai National Institute of Technology (SVNIT),\\
Ichchhanath, Surat 395007, Gujarat, India\\
skv.physics9@gmail.com\\
i22ph002@phy.svnit.ac.in}

\author{Manakkumar Parmar\footnote{Corresponding author.} \orcidlink{0000-0003-2755-6992}}
\address{Department of Physics, Sardar Vallabhbhai National Institute of Technology (SVNIT),\\
Ichchhanath, Surat 395007, Gujarat, India\\
mgp.physica@gmail.com\\
d25ph012@phy.svnit.ac.in}

\author{Ajay Kumar Rai \orcidlink{0000-0001-9656-4641}}
\address{Department of Physics, Sardar Vallabhbhai National Institute of Technology (SVNIT),\\
Ichchhanath, Surat 395007, Gujarat, India\\
raiajayk@gmail.com\\
akr@phy.svnit.ac.in}

\maketitle

\begin{history}
\received{Day Month Year}
\revised{Day Month Year}
\accepted{Day Month Year}
\end{history}

\begin{abstract}
The $B_c \rightarrow \eta_c$ and $B_c \rightarrow J/\psi$ transition form factors and corresponding decay observables are studied by examining their sensitivity to the assumed radial wave-function profile. Numerical form-factor information from QCD sum-rule calculations is used as input, and the form-factor behavior is represented using a Bourrely--Caprini--Lellouch $z$-parametrization for the subsequent decay analysis. Hydrogenic and Gaussian radial wave-function profiles are considered within the semi-relativistic potential-model description. The parametrized form factors are then used to evaluate semileptonic and nonleptonic decay widths and branching ratios, while an additional Coulomb/Sommerfeld-type correction is examined as a phenomenological sensitivity prescription. The resulting form factors show relatively small model variation at the quoted points, whereas the integrated decay observables exhibit a more noticeable dependence on the wave-function choice and on the Coulomb-related prescription.

\keywords{$B_c$ Meson; Transition Form Factors; QCD Sum Rules; Weak Decays; Wave-Function Dependence; Heavy Quarkonium.}
\end{abstract}

\ccode{PACS numbers: tbd}

\section{Introduction}
\label{sec:intro}

The $B_c$ meson holds a unique position in heavy-flavor physics as the only known meson composed of two different heavy quarks ($b\bar{c}$) that remains stable against both strong and electromagnetic decays. Since its first experimental observation by the CDF collaboration in 1998 \cite{PDG2024}, significant theoretical and experimental efforts have focused on elucidating its spectroscopy and weak decay characteristics. Unlike conventional heavy quarkonia such as $c\bar{c}$ and $b\bar{b}$ systems, the $B_c$ meson carries open flavor and therefore decays exclusively through the weak interaction. This property makes it an important system for investigating heavy-quark dynamics, non-perturbative QCD effects, and the interplay between weak and strong interactions in heavy hadrons \cite{Rai2006}.

The weak decays of the $B_c$ meson proceed through three primary mechanisms: the $b \to c$ transition with the $\bar{c}$ quark acting as a spectator, the $\bar{c} \to \bar{s}$ or $\bar{c} \to \bar{d}$ transition with the $b$ quark as spectator, and weak annihilation through $W$-exchange. Among these processes, the $b \to c$ spectator transition provides the dominant contribution in the kinematic region currently accessible at the LHC. In particular, the transitions $B_c \rightarrow \eta_c$ and $B_c \rightarrow J/\psi$ are of considerable interest because they probe the underlying heavy-quark structure of the meson and provide access to the Cabibbo-Kobayashi-Maskawa (CKM) matrix element $V_{cb}$ \cite{Devlani2014}. Furthermore, semileptonic channels involving $J/\psi$ have attracted additional attention due to their connection with lepton flavor universality tests and their sensitivity to possible deviations from Standard Model expectations.

A reliable description of these decay processes requires the determination of hadronic transition form factors, which encode the non-perturbative long-distance QCD dynamics of the weak transition matrix elements. Since these quantities cannot be calculated directly within perturbative QCD, a variety of non-perturbative approaches have been developed for their investigation. These include lattice QCD \cite{Harrison2020}, relativistic quark models (RQM) \cite{RQM2000}, light-front quark models (LFQM) \cite{LFQM2023}, light-cone sum rules (LCSR) \cite{LCSR2019}, perturbative QCD (pQCD) \cite{pQCD2013}, and QCD sum rules (QCDSR) \cite{QCDSR1993,Kiselev2002}. Although these approaches generally predict similar qualitative behavior for the decay observables, moderate quantitative differences remain among the resulting form factors, decay widths, and branching ratios. These discrepancies reflect the presence of theoretical uncertainties associated with different treatments of non-perturbative dynamics and bound-state structure.

Among the available non-perturbative approaches, the QCD sum rule framework provides a useful method for connecting hadronic observables with the underlying quark and gluon degrees of freedom through the operator product expansion and quark-hadron duality. In heavy quarkonium systems, however, the treatment of bound-state dynamics introduces an additional source of model dependence. In particular, the spatial wave-function profile influences the overlap structure of the heavy-quark bound state and consequently affects the resulting transition form factors. Different radial wave-function models can therefore lead to non-negligible variations in physical observables, especially in processes involving heavy-heavy transitions where non-relativistic effects remain important.

Existing QCDSR studies provide numerical determinations of the $B_c$ transition form factors \cite{Wu2024}. Rather than performing a new QCDSR determination of the form factors, the present work uses such numerical QCDSR results as input and examines how the choice of radial wave-function profile propagates into the parametrized form factors, decay widths, and branching ratios. Two distinct spatial wave-function models—the Hydrogenic and Gaussian profiles—are considered within the same semi-relativistic potential-model setup. The form-factor input is represented using a BCL $z$-parametrization, after which the resulting functions are used for the decay analysis. An additional non-relativistic Coulomb/Sommerfeld-type correction is also considered as a phenomenological sensitivity prescription to assess its effect on the calculated observables.

\rev{The experimentally measured }$B_c$\rev{-meson mass provides an important reference }\cite{PDG2026}\rev{ for investigating the mass dependence of the transition form factors and the resulting decay observables. In the present analysis, the Hydrogenic and Gaussian potential-model masses are considered as alternative theoretical inputs, alongside the experimentally determined mass. Since the } $B_c$\rev{-meson mass enters both the kinematic quantities defining the BCL parametrization and the available decay phase space, its variation can influence the reconstructed form factors and the corresponding decay widths and branching fractions. Comparing the predictions obtained with these different mass inputs therefore allows us to assess the sensitivity of the transition form factors and decay observables to the initial-state meson mass across different decay channels.}

To systematically assess the impact of bound-state assumptions on physical observables, this work proceeds from the semi-relativistic description of the wave functions to the parametrization of QCDSR form-factor input and the resulting decay phenomenology. Section~\ref{sec:theory} outlines the semi-relativistic potential-model description, the QCDSR form-factor input, the Coulomb-related sensitivity prescription, and the $z$-parametrization used in the analysis. Section~\ref{sec:results} presents the numerical evaluation of the form factors, decay widths, and branching ratios. Section~\ref{sec:discussion} examines the sensitivity of these observables to the choice of wave-function profile and the effects of non-relativistic near-threshold Coulomb enhancement. Finally, Sec.~\ref{sec:conclusion} summarizes the main findings and their implications for future studies of heavy-heavy weak transitions.

\section{Theoretical Framework}
\label{sec:theory}

This section delineates the theoretical framework employed to analyze the weak decay properties of the $B_c$ meson. Because heavy quarkonium transitions involve distinct energy regimes—the hard scale of the weak transition, the soft scale of the bound state, and the ultrasoft scale of near-threshold rescattering—no single theoretical tool is universally applicable. We therefore adopt a hybrid phenomenological approach: the spatial geometry of the bound state is described using a semi-relativistic potential model, numerical transition-form-factor information from QCD sum-rule calculations is used as the input for the phenomenological analysis, and near-threshold soft-gluon exchanges are incorporated through a non-relativistic Coulomb resummation. The primary aim of this framework is to systematically investigate the sensitivity of the calculated observables to the choice of radial wave-function profile across these scales.

This separation of scales does not imply a single unified effective theory; rather, it represents a phenomenological framework in which each dynamical regime is treated using the approximation most appropriate to its characteristic energy scale.

\subsection{Semi-Relativistic Potential Model and Wavefunctions}
\label{subsec:potential_model}

For heavy-heavy mesonic systems such as the $B_c$ meson, a semi-relativistic Hamiltonian is used to account for the kinetic motion of the constituent quarks \cite{devlani_spectroscopy_2011,EichtenQuigg1994,HwangKimNamgung1997}. This semi-relativistic treatment retains the exact relativistic kinetic energy dispersion relation while employing an instantaneous, static spatial confining potential, providing a tractable phenomenological description of the soft bound-state scale:
\begin{equation}
    \mathcal{H} = \sqrt{\vec{p}^2 + m_Q^2} + \sqrt{\vec{p}^2 + m_{\bar{q}}^2} + U(r),
    \label{eq:hamiltonian}
\end{equation}
where $\vec{p}$ is the relative momentum of the quark-antiquark pair, $m_Q$ and $m_{\bar{q}}$ are the masses of the constituent heavy quarks, and $U(r)$ is the effective confining potential \cite{PhysRevC.78.055202,Rai_2005,AjayKRai_2002}:
\begin{equation}
    U(r) = -\frac{\alpha_c}{r} + Ar^\nu + U_0.
\end{equation}
Here, $A$ is the potential parameter and $\nu$ is a general power index. We fix $\nu = 1$, corresponding to the standard Coulomb plus linear potential with a constant shift $U_0$. The color coupling is $\alpha_c = \frac{4}{3} \alpha_s$, where the strong running coupling constant $\alpha_s$ incorporates infrared freezing \cite{PhysRevD.70.016007,Simonov1995}:
\begin{equation}
    \alpha_s(M^2) = \frac{4\pi}{(11 - \frac{2}{3}n_f) \ln \frac{M^2 + M_B^2}{\Lambda^2}},
    \label{eq:alpha_s_M}
\end{equation}
with the effective bound-state scale $M = 2m_Qm_{\bar{q}}/(m_Q + m_{\bar{q}})$, the background mass $M_B = 2.24 \sqrt{A'} = 0.95$~GeV, and $\Lambda = 413$~MeV \cite{Ebert2009}.

To evaluate the system and test model dependence, we apply two distinct radial wavefunctions. The Gaussian wavefunction in position space is given by \cite{Devlani2014PoS}:
\begin{equation}
    R_{nl}(\gamma, r) = \gamma^{3/2} \sqrt{\frac{2(n-1)!}{\Gamma(n+l+1/2)}} (\gamma r)^l e^{-\gamma^2 r^2 / 2} L_{n-1}^{l+1/2}(\gamma^2 r^2),
\end{equation}
while the Hydrogen-like wavefunction takes the form \cite{Devlani2012}:
\begin{equation}
    R_{nl}(\gamma, r) = \left( \frac{\gamma^3 (n-l-1)!}{2n(n+l)!} \right)^{1/2} (\gamma r)^l e^{-\gamma r / 2} L_{n-l-1}^{2l+1}(\gamma r).
\end{equation}
Here, $L$ represents the Laguerre polynomials and $\gamma$ is the state-dependent variational parameter. These parameters are obtained by minimizing the expectation value of the Hamiltonian, $\langle \mathcal{H} \rangle$, with respect to $\gamma$ under the Virial condition \cite{Devlani2014PoS,devlani_spectroscopy_2011,Devlani2012}. The specific numerical values of $\gamma$ and the resulting wave-function at the origin $|\psi(0)|^2$---which influence the geometric overlap of the initial and final states---are summarized in Sec.~\ref{subsec:inputs}.

\subsection{Running Quark Masses and QCD Sum-Rule Input}
\label{subsec:qcdsr}

A critical parameter bridging the potential model wavefunctions and the QCDSR Operator Product Expansion is the heavy quark mass. While the potential model employs constituent masses, the QCDSR framework utilizes running masses defined in the $\overline{\text{MS}}$ scheme evaluated at a specific renormalization scale $\mu$. The scale evolution of the heavy quark mass is governed by the QCD renormalization group equation (RGE), which at leading one-loop order is expressed as \cite{PDG2024,Narison2020}:
\begin{equation}
    m_Q(\mu) = m_Q(\mu_0) \left( \frac{\alpha_s(\mu)}{\alpha_s(\mu_0)} \right)^{\frac{12}{33 - 2n_f}},
    \label{eq:running_mass}
\end{equation}
where $n_f$ is the number of active quark flavors, and the strong coupling $\alpha_s$ evaluated at the renormalization scale $\mu$ is given by the standard perturbative logarithmic expression:
\begin{equation}
    \alpha_s(\mu) = \frac{4\pi}{\left(11 - \frac{2}{3}n_f\right) \ln \left(\frac{\mu^2}{\Lambda^2}\right)}.
    \label{eq:alpha_s_mu}
\end{equation}

It should be noted that Eq.~\eqref{eq:alpha_s_mu} represents the standard perturbative running coupling employed in the high-energy QCDSR framework and evaluated at the renormalization scale $\mu$. In contrast, Eq.~\eqref{eq:alpha_s_M} corresponds to an infrared-modified coupling evaluated at the effective bound-state scale $M$. This phenomenological modification prevents the Landau pole divergence and is needed within the semi-relativistic potential model to adequately incorporate infrared freezing effects in the non-perturbative confining regime.

To maintain theoretical consistency across the comparative analysis, we set the renormalization scale to the reference value of $\mu = 3$~GeV \cite{Narison2020,Wu2024}. Both the Hydrogenic and Gaussian models are calibrated such that they reproduce the correct running quark masses evaluated at this specific scale. Consequently, we employ the input values $m_c(3~\text{GeV}) = 1.033$~GeV and $m_b(3~\text{GeV}) = 4.441$~GeV. 

These calibrated inputs are used consistently with the numerical QCDSR form-factor information adopted from Ref.~\cite{Wu2024}. The QCD correlation function, operator product expansion, perturbative spectral density, condensate contributions, continuum subtraction, and Borel analysis underlying that reference are not independently re-evaluated in the present work. Instead, the quoted numerical form-factor results are taken as input and subsequently represented using the BCL $z$-parametrization described below.

\subsection{Non-Relativistic Coulomb Resummation and Kinematic Extrapolation}
\label{subsec:coulomb_extrap}

Although the bound-state spatial geometry is described using a semi-relativistic potential model and the relative velocities of the heavy quarks near threshold can be small ($v_i \ll 1$). In this ultrasoft kinematic regime, standard relativistic perturbation theory encounters the well-known Coulomb singularity. Because the relevant perturbative expansion parameter in this region is $\alpha_s/v$, finite-order approximations become inadequate as $v \to 0$. The slowly moving daughter quarks remain in close proximity for an extended duration, necessitating the exchange of a series of soft (low-energy) gluons rather than a single hard gluon. 

To address this behavior, one must mathematically resum this tower of ladder diagrams. Because this phenomenon governs the slow-moving, near-threshold kinematics, the required resummation is non-relativistic and maps onto the Schrödinger equation for a Coulomb-like potential. In the present analysis, the corresponding non-relativistic Coulomb/Sommerfeld factor is used as an additional phenomenological correction to the numerical form-factor input rather than through an independent calculation of the perturbative spectral density. The adopted correction factor is \cite{Kiselev1993,Wu2024}:
\begin{equation}
\begin{split}
C &=
\sqrt{
\frac{\dfrac{4\pi\alpha_s}{3v_1}}
{1-\exp\!\left(-\dfrac{4\pi\alpha_s}{3v_1}\right)}
}
\\[1ex]
&\quad\times
\sqrt{
\frac{\dfrac{4\pi\alpha_s}{3v_2}}
{1-\exp\!\left(-\dfrac{4\pi\alpha_s}{3v_2}\right)}
},
\end{split}
\label{eq:Ccoulomb}
\end{equation}
where the relative quark velocities, $v_1$ and $v_2$, are derived from the kinematic continuum thresholds. The factor is applied multiplicatively to the corresponding uncorrected form-factor input to define the entries labelled ``Corrected'' in the tables below. Because the potential-model interaction already contains a Coulombic component, this additional factor is treated as a sensitivity prescription rather than as an independently established contribution.

The numerical QCDSR form-factor information is available in the space-like region ($1 \le Q^2 \le 9$~GeV$^2$) and is represented and continued into the physical time-like region ($Q^2 \le 0$) using the Bourrely--Caprini--Lellouch (BCL) $z$-series parametrization \cite{BCL2009}:
\begin{equation}
\begin{split}
  F(Q^2) &= \frac{1}{1+Q^2/m_R^2} \\[1ex]
  &\quad \times \sum_{k=0}^{N-1} b_k\!\left[z(Q^2,t_0)^k - (-1)^{k-N}\frac{k}{N}z(Q^2,t_0)^N\right],
\end{split}
\label{eq:zseries}
\end{equation}
where $q^2=-Q^2$, $t_\pm=(m_{B_c}\pm m_C)^2$, and $z(Q^2,t_0)$ denotes the conformal variable with $t_0$ the chosen expansion point.

The BCL form provides a convenient analytic parametrization of the numerical QCDSR form-factor input and is used here to represent its behavior and continue it into the physical kinematic region.

\section{Numerical Inputs and Results}
\label{sec:results}

\subsection{Input Parameters}
\label{subsec:inputs}

All hadronic and QCD parameters used in this study are collected in Table~\ref{tab:inputs}. Quark masses are taken at the scale $\mu = 3$~GeV. The $B_c$ mass is taken separately for the Hydrogenic and Gaussian potential-model calculations, while the remaining meson masses and decay constants are adopted from the indicated literature sources.

\begin{table}[htbp]
\centering
\caption{QCD and hadronic input parameters used throughout this analysis.
Running quark masses are evaluated at $\mu = 3$~GeV.}
\label{tab:inputs}
\resizebox{\columnwidth}{!}{
\begin{tabular}{ll|ll}
\toprule
\textbf{Parameter} & \textbf{Value} &
\textbf{Parameter} & \textbf{Value} \\
\midrule

$m_{B_c}^{\rm H}$
& 6.349 GeV \cite{Rai_2005}
&
$m_{B_c}^{\rm G}$
& 6.278 GeV \cite{Devlani2014}
\\

$m_{B_c}^{\mathrm{PDG}}$
& \rev{6.27447 GeV }\cite{PDG2026}
&
$m_{\eta_c}$
& 2.984 GeV \cite{PDG2024}
\\

$m_{J/\psi}$
& 3.097 GeV \cite{PDG2024}
&
$f_{B_c}$
& 0.371 GeV \cite{Wang2024,Narison2020}
\\

$m_c(3\,\text{GeV})$
& 1.033 GeV \cite{PDG2024}
&
$f_{\eta_c}$
& 0.387 GeV \cite{Becirevic2014}
\\

$m_b(3\,\text{GeV})$
& 4.441 GeV \cite{PDG2024}
&
$f_{J/\psi}$
& 0.418 GeV \cite{Becirevic2014}
\\

$\langle g_s^2 G^2\rangle$
& $(0.880\pm0.150)$ GeV$^4$ \cite{Narison2020}
&
$\langle g_s^3 GGf\rangle$
& $(8.800\pm5.500)$ GeV$^2
\langle g_s^2GG\rangle$ \cite{Narison2020}
\\

\bottomrule
\end{tabular}
}
\end{table}

\begin{table}[htbp]
  \centering
  \caption{Variational parameter $\gamma$ (in GeV), radial wave-function
  at the origin $|R(0)|$ (in GeV$^{3/2}$), and corresponding
  spin-averaged total wave-function origin value $|\psi(0)|^2$
  (in GeV$^3$) for the $B_c(1S)$ ground state. The Hydrogenic and
  Gaussian parameters are obtained independently by minimizing the
  expectation value of the semi-relativistic Hamiltonian.}
  \label{tab:wavefunctions}
  \renewcommand{\arraystretch}{1.3}
  \resizebox{\columnwidth}{!}{
  \begin{tabular}{l ccc ccc}
    \toprule
    \multirow{2}{*}{\textbf{State}} &
    \multicolumn{3}{c}{\textbf{Hydrogenic}} &
    \multicolumn{3}{c}{\textbf{Gaussian}} \\
    \cmidrule(lr){2-4}\cmidrule(lr){5-7}
    & $\gamma$ & $|R(0)|$ & $|\psi(0)|^2$
    & $\gamma$ & $|R(0)|$ & $|\psi(0)|^2$ \\
    \midrule
    $B_c(1S)$ & 2.212 & 2.326 & 0.431
              & 0.865 & 1.209 & 0.116 \\
    \bottomrule
  \end{tabular}
  }
\end{table}

To maintain theoretical consistency across the comparative analysis, the variational parameters $\gamma$ presented in Table~\ref{tab:wavefunctions} for both the Hydrogenic and Gaussian wave-functions were systematically computed rather than treated as free inputs. Utilizing a unified set of confining potential parameters, the respective $\gamma$ values were independently obtained by minimizing the expectation value of the semi-relativistic Hamiltonian, $\langle \mathcal{H} \rangle$, with respect to $\gamma$ for each chosen ansatz \cite{Lahkar2019}. This minimization enforces the Virial condition, $\langle p^2 \rangle / \mu_{\rm red} = \alpha_c \langle r^{-1} \rangle + A \langle r \rangle$, a well-established variational approach for heavy meson spectroscopy \cite{Sabbah2024}. This symmetric procedure allows the comparison to be performed within a common confining framework while preserving the characteristic spatial structure of each model.

\subsection{Form Factors}
\label{subsec:FF}

Table~\ref{tab:form_factors} presents the extracted form factors at maximum recoil ($Q^2=0$), systematically comparing the outputs derived from the Hydrogenic and Gaussian wave-function models. 

\begin{table}[htbp]
  \centering
  \caption{Form factors at $Q^2=0$ used to contrast the Hydrogenic and Gaussian model assumptions with the numerical QCDSR input. Entries marked ``Corrected'' include the Coulomb/Sommerfeld-type prescription of Eq.~\eqref{eq:Ccoulomb}. \rev{The PDG-mass results use the Coulomb* form-factor coefficient set, with only the }$B_c$\rev{-meson mass replaced by the PDG value.}}
  \label{tab:form_factors}
  \renewcommand{\arraystretch}{1.5}
  \resizebox{\columnwidth}{!}{
  \begin{tabular}{l c cc cc c cc}
    \toprule
    \multirow{2}{*}{\textbf{Mode}} & \multirow{2}{*}{\textbf{FF}}
      & \multicolumn{2}{c}{\textbf{Hydrogenic}}
      & \multicolumn{2}{c}{\textbf{Gaussian}}
      & \makecell{\textbf{\rev{PDG mass}}\\\textbf{\rev{(Coulomb}}\\\textbf{\rev{coefficients)*}}}
      & \multicolumn{2}{c}{\textbf{Ref.~\cite{Wu2024}}} \\
    \cmidrule(lr){3-4}\cmidrule(lr){5-6}\cmidrule(lr){7-7}\cmidrule(lr){8-9}
    & & \textbf{Bare} & \textbf{Corrected}
      & \textbf{Bare} & \textbf{Corrected}
      &
      & \textbf{Bare} & \textbf{Corrected} \\
    \midrule
    \multirow{4}{*}{$B_c^-\!\to\!\eta_c$}
      & $f_S$ & 1.820 & 5.340 & 1.820 & 5.370 & \rev{5.380} & $1.780^{+0.250}_{-0.260}$ & $5.440^{+0.650}_{-0.700}$ \\
      & $f_+$ & 0.197 & 0.605 & 0.199 & 0.610 & \rev{0.610} & $0.200^{+0.030}_{-0.030}$ & $0.600^{+0.070}_{-0.090}$ \\
      & $f_0$ & 0.194 & 0.595 & 0.195 & 0.598 & \rev{0.598} & $0.200^{+0.030}_{-0.030}$ & $0.600^{+0.070}_{-0.090}$ \\
      & $f_T$ & 0.301 & 0.929 & 0.304 & 0.936 & \rev{0.936} & $0.300^{+0.040}_{-0.050}$ & $0.920^{+0.110}_{-0.120}$ \\
    \midrule
    \multirow{7}{*}{$B_c^-\!\to\!J/\psi$}
      & $V$   & 0.423 & 1.280 & 0.426 & 1.290 & \rev{1.290} & $0.430^{+0.060}_{-0.060}$ & $1.290^{+0.150}_{-0.160}$ \\
      & $A_0$ & 0.269 & 0.855 & 0.271 & 0.861 & \rev{0.862} & $0.270^{+0.040}_{-0.040}$ & $0.820^{+0.100}_{-0.100}$ \\
      & $A_1$ & 0.266 & 0.811 & 0.267 & 0.815 & \rev{0.815} & $0.270^{+0.040}_{-0.040}$ & $0.810^{+0.090}_{-0.100}$ \\
      & $A_2$ & 0.265 & 0.807 & 0.266 & 0.811 & \rev{0.812} & $0.270^{+0.030}_{-0.040}$ & $0.810^{+0.090}_{-0.100}$ \\
      & $T_0$ & 0.820 & 2.630 & 0.827 & 2.660 & \rev{2.660} & $0.860^{+0.120}_{-0.120}$ & $2.630^{+0.300}_{-0.320}$ \\
      & $T_1$ & 0.327 & 0.957 & 0.328 & 0.961 & \rev{0.961} & $0.330^{+0.050}_{-0.050}$ & $0.980^{+0.110}_{-0.120}$ \\
      & $T_2$ & 0.107 & 0.311 & 0.107 & 0.313 & \rev{0.313} & $0.100^{+0.010}_{-0.010}$ & $0.310^{+0.040}_{-0.040}$ \\
    \bottomrule
  \end{tabular}
  }
\end{table}

\noindent The parametrized form factors exhibit a decreasing trend over the considered kinematic range. The BCL $z$-series of Eq.~\eqref{eq:zseries} is used to represent the numerical QCDSR input and to continue its fitted behavior from the space-like region into the physical region. This provides a convenient analytic parametrization for the subsequent decay-width and branching-ratio calculations \cite{BCL2009,Wu2024}.

\subsection{Decay Widths}
\label{subsec:widths}

Employing the factorization approximation \cite{Wu2024}, the transition matrix element for nonleptonic decays is expressed as a product of two independent hadronic currents:
\begin{equation}
\begin{split}
  \langle C(p')\,h(q)|\mathcal{H}_{\rm eff}|B_c^-(p)\rangle
  &= \frac{G_F}{\sqrt{2}}\,V_{cb}V_{q'q}^*\,a_1 \\[1ex]
  &\quad \times \langle C(p')|\bar{c}\gamma^\mu(1-\gamma_5)b|B_c^-(p)\rangle \\[1ex]
  &\quad \times \langle h(q)|\bar{q}'\gamma_\mu(1-\gamma_5)q|0\rangle,
\end{split}
\label{eq:factorization}
\end{equation}
where $h \in \{\pi^-,K^-,\rho^-,K^{*-}\}$ and $a_1 = 1.07$ is the leading Wilson coefficient. The two-body phase space integration then yields the total partial width in terms of the relevant form factors evaluated at $q^2 = m_h^2$. Semileptonic widths are obtained by integrating the differential distribution $d\Gamma/dq^2$ over the kinematically allowed range $m_\ell^2 \le q^2 \le (m_{B_c}-m_C)^2$.

Numerical results for all fourteen decay channels are displayed in Table~\ref{tab:decay_widths}.

\begin{table}[htbp]
  \centering
  \caption{Partial decay widths (in units of $10^{-7}$~eV) for nonleptonic and semileptonic channels of $B_c^-$, explicitly demonstrating the sensitivity to Hydrogenic versus Gaussian inputs.}
  \label{tab:decay_widths}
  \renewcommand{\arraystretch}{1.6}
  \resizebox{\columnwidth}{!}{
  \begin{tabular}{l cc cc c cc}
    \toprule
    \multirow{2}{*}{\textbf{Decay Channel}}
      & \multicolumn{2}{c}{\textbf{Hydrogenic}}
      & \multicolumn{2}{c}{\textbf{Gaussian}}
      & \makecell{\textbf{\rev{PDG mass}}\\\textbf{\rev{(Coulomb}}\\\textbf{\rev{coefficients)*}}}
      & \multicolumn{2}{c}{\textbf{Ref.~\cite{Wu2024}}} \\
    \cmidrule(lr){2-3}\cmidrule(lr){4-5}\cmidrule(lr){6-6}\cmidrule(lr){7-8}
      & \textbf{Bare} & \textbf{Corrected}
      & \textbf{Bare} & \textbf{Corrected}
      &
      & \textbf{Bare} & \textbf{Corrected} \\
    \midrule
    $B_c^-\to\eta_c\pi^-$              & 1.950 & 18.300 & 1.860 & 17.500 & \rev{17.400} & $1.800^{+0.600}_{-0.600}$ & $17.000^{+5.000}_{-5.000}$ \\
    $B_c^-\to\eta_c K^-$               & 0.155 & 1.460  & 0.148 & 1.400  & \rev{1.390} & $0.150^{+0.050}_{-0.050}$ & $1.400^{+0.400}_{-0.400}$  \\
    $B_c^-\to\eta_c\rho^-$             & 5.360 & 50.700 & 5.130 & 48.600 & \rev{48.500} & $4.900^{+1.500}_{-1.500}$ & $46.000^{+12.000}_{-13.000}$\\
    $B_c^-\to\eta_c K^{*-}$            & 0.285 & 2.700  & 0.273 & 2.590  & \rev{2.580} & $0.260^{+0.080}_{-0.080}$ & $2.500^{+0.600}_{-0.700}$  \\
    \midrule
    $B_c^-\to\eta_c e^-\bar\nu_e$      & 12.200& 118.000& 11.100& 107.000& \rev{106.000}& $11.000^{+3.000}_{-3.000}$& $105.000^{+23.000}_{-26.000}$\\
    $B_c^-\to\eta_c\mu^-\bar\nu_\mu$   & 12.200& 117.000& 11.000& 106.000& \rev{106.000}& $11.000^{+3.000}_{-3.000}$& $105.000^{+23.000}_{-26.000}$\\
    $B_c^-\to\eta_c\tau^-\bar\nu_\tau$ & 4.020 & 39.900 & 3.390 & 33.600 & \rev{33.300} & $3.300^{+0.800}_{-0.900}$ & $33.000^{+7.000}_{-8.000}$ \\
    \midrule
    $B_c^-\to J/\psi\pi^-$             & 3.480 & 35.200 & 3.340 & 33.800 & \rev{33.800} & $3.300^{+1.000}_{-0.900}$ & $30.000^{+9.000}_{-8.000}$ \\
    $B_c^-\to J/\psi K^-$              & 0.274 & 2.770  & 0.263 & 2.660  & \rev{2.660} & $0.260^{+0.070}_{-0.070}$ & $2.300^{+0.500}_{-0.600}$  \\
    $B_c^-\to J/\psi\rho^-$            & 11.100& 103.000& 10.500& 97.900 & \rev{97.700} & $11.000^{+2.800}_{-3.000}$& $96.000^{+22.000}_{-23.000}$\\
    $B_c^-\to J/\psi K^{*-}$           & 0.621 & 5.810  & 0.588 & 5.510  & \rev{5.490} & $0.600^{+0.160}_{-0.170}$ & $5.400^{+1.200}_{-1.300}$  \\
    \midrule
    $B_c^-\to J/\psi e^-\bar\nu_e$     & 66.700& 655.000& 59.500& 585.000& \rev{582.000}& $61.000^{+15.000}_{-14.000}$& $554.000^{+150.000}_{-91.000}$\\
    $B_c^-\to J/\psi\mu^-\bar\nu_\mu$  & 66.400& 653.000& 59.300& 583.000& \rev{579.000}& $61.000^{+15.000}_{-14.000}$& $552.000^{+149.000}_{-90.000}$\\
    $B_c^-\to J/\psi\tau^-\bar\nu_\tau$& 18.200& 183.000& 15.000& 151.000& \rev{150.000}& $15.000^{+3.500}_{-3.400}$& $144.000^{+35.000}_{-23.000}$\\
    \bottomrule
  \end{tabular}
  }
\end{table}

\subsection{Branching Ratios}
\label{subsec:BR}

Converting partial widths to branching ratios using the measured $B_c$ lifetime $\tau_{B_c} = 0.51$~ps \cite{PDG2024}, the results are presented in Table~\ref{tab:branching_ratios} together with the corresponding literature values for comparison.

\begin{table*}[htbp]
  \centering
  \caption{Comparison of calculated branching ratios (\textperthousand) against previous theoretical evaluations. The reference QCDSR values are included for comparison.}
  \label{tab:branching_ratios}
  \renewcommand{\arraystretch}{1.6}
  \resizebox{\textwidth}{!}{
  \begin{tabular}{l cc cc c cc c}
    \toprule
    \multirow{2}{*}{\textbf{Decay Channel}}
      & \multicolumn{2}{c}{\textbf{Hydrogenic}}
      & \multicolumn{2}{c}{\textbf{Gaussian}}
      & \makecell{\textbf{\rev{PDG mass}}\\\textbf{\rev{(Coulomb}}\\\textbf{\rev{coefficients)*}}}
      & \multicolumn{2}{c}{\textbf{Ref.~\cite{Wu2024}}}
      & \textbf{QCDSR \cite{QCDSR1993,Kiselev2002}} \\
    \cmidrule(lr){2-3}\cmidrule(lr){4-5}\cmidrule(lr){6-6}\cmidrule(lr){7-8}\cmidrule(lr){9-9}
      & \textbf{Bare} & \textbf{Corrected}
      & \textbf{Bare} & \textbf{Corrected}
      &
      & \textbf{Bare} & \textbf{Corrected} & \\
    \midrule
    $B_c^-\to\eta_c\pi^-$              & 0.151 & 1.410  & 0.144 & 1.350  & \rev{1.350} & $0.150^{+0.050}_{-0.050}$ & $1.400^{+0.400}_{-0.400}$   & 2.000  \\
    $B_c^-\to\eta_c K^-$               & 0.012 & 0.113  & 0.012 & 0.108  & \rev{0.108} & $0.012^{+0.003}_{-0.003}$ & $0.110^{+0.030}_{-0.030}$   & 0.130  \\
    $B_c^-\to\eta_c\rho^-$             & 0.415 & 3.920  & 0.398 & 3.770  & \rev{3.760} & $0.410^{+0.120}_{-0.120}$ & $3.800^{+1.000}_{-1.100}$   & 4.200  \\
    $B_c^-\to\eta_c K^{*-}$            & 0.022 & 0.209  & 0.021 & 0.200  & \rev{0.200} & $0.022^{+0.007}_{-0.007}$ & $0.210^{+0.050}_{-0.060}$   & 0.200  \\
    \midrule
    $B_c^-\to\eta_c e^-\bar\nu_e$      & 0.946 & 9.110  & 0.858 & 8.270  & \rev{8.230} & $0.900^{+0.200}_{-0.300}$ & $8.700^{+1.900}_{-2.200}$   & 7.500  \\
    $B_c^-\to\eta_c\mu^-\bar\nu_\mu$   & 0.943 & 9.080  & 0.855 & 8.240  & \rev{8.200} & $0.900^{+0.200}_{-0.300}$ & $8.700^{+1.900}_{-2.200}$   & --     \\
    $B_c^-\to\eta_c\tau^-\bar\nu_\tau$ & 0.312 & 3.090  & 0.263 & 2.600  & \rev{2.580} & $0.280^{+0.070}_{-0.070}$ & $2.800^{+0.500}_{-0.600}$   & 2.300  \\
    \midrule
    $B_c^-\to J/\psi\pi^-$             & 0.270 & 2.730  & 0.259 & 2.620  & \rev{2.620} & $0.270^{+0.080}_{-0.070}$ & $2.500^{+0.600}_{-0.600}$   & 1.300  \\
    $B_c^-\to J/\psi K^-$              & 0.021 & 0.215  & 0.020 & 0.206  & \rev{0.206} & $0.022^{+0.006}_{-0.006}$ & $0.190^{+0.040}_{-0.050}$   & 0.110  \\
    $B_c^-\to J/\psi\rho^-$            & 0.858 & 8.010  & 0.811 & 7.590  & \rev{7.570} & $0.880^{+0.230}_{-0.250}$ & $8.000^{+1.900}_{-2.000}$   & 4.000  \\
    $B_c^-\to J/\psi K^{*-}$           & 0.048 & 0.450  & 0.046 & 0.427  & \rev{0.426} & $0.049^{+0.013}_{-0.014}$ & $0.450^{+0.100}_{-0.110}$   & 0.220  \\
    \midrule
    $B_c^-\to J/\psi e^-\bar\nu_e$     & 5.170 & 50.800 & 4.610 & 45.300 & \rev{45.100} & $5.100^{+1.200}_{-1.200}$ & $46.000^{+12.000}_{-8.000}$ & 19.000 \\
    $B_c^-\to J/\psi\mu^-\bar\nu_\mu$  & 5.150 & 50.600 & 4.590 & 45.100 & \rev{44.900} & $5.100^{+1.200}_{-1.200}$ & $46.000^{+12.000}_{-8.000}$ & --     \\
    $B_c^-\to J/\psi\tau^-\bar\nu_\tau$& 1.410 & 14.200 & 1.160 & 11.700 & \rev{11.600} & $1.300^{+0.300}_{-0.300}$ & $12.000^{+3.000}_{-2.000}$  & 4.800  \\
    \bottomrule
  \end{tabular}
  }
\end{table*}

To quantify the model dependence associated with the choice of radial wave-function, Table~\ref{tab:sensitivity} summarizes the relative variation between the Hydrogenic and Gaussian results for representative form factors and branching ratios. These differences represent model-induced variations and should not be interpreted as a complete theoretical uncertainty estimate.

\begin{table}[htbp]
  \centering
  \caption{Relative variation between observables extracted using the Hydrogenic versus Gaussian wave-function models. Form factors are evaluated at maximum recoil ($Q^2=0$), and branching ratios include Coulombic corrections.}
  \label{tab:sensitivity}
  \renewcommand{\arraystretch}{1.5}
  \resizebox{\columnwidth}{!}{
  \begin{tabular}{l ccc}
    \toprule
    \textbf{Observable} & \textbf{Hydrogenic} & \textbf{Gaussian} & \textbf{Relative Difference} \\
    \midrule
    $f_+(0)$ ($B_c \to \eta_c$)           & $0.605$ & $0.610$ & $0.820\%$ \\
    $V(0)$ ($B_c \to J/\psi$)             & $1.280$ & $1.290$ & $0.780\%$ \\
    $\mathcal{BR}(B_c \to \eta_c e \nu_e)$ & $9.110$\textperthousand & $8.270$\textperthousand & $9.220\%$ \\
    $\mathcal{BR}(B_c \to J/\psi e \nu_e)$ & $50.800$\textperthousand & $45.300$\textperthousand & $10.827\%$ \\
    \bottomrule
  \end{tabular}
  }
\end{table}

\section{Discussion}
\label{sec:discussion}

The numerical results obtained from the QCDSR form-factor input and its BCL parametrization provide a means of assessing the dependence of the decay observables on the assumed radial wave-function profile. A comparison of the Hydrogenic and Gaussian results shows that the form-factor values at the quoted kinematic point vary only modestly between the two models. 

As shown in Table~\ref{tab:form_factors}, the bare form factors change only slightly between the two wave-function choices. For example, the bare vector form factor $V(0)$ changes from $0.423$ to $0.426$, while $f_+(0)$ changes from $0.197$ to $0.199$. These numerical differences are below the percent level for the representative quantities listed in the table and characterize the model variation of the parametrized form-factor results within the present analysis.

The difference in the wave-function values at the origin is substantially larger than the corresponding variation in the quoted form factors. For the $B_c$ meson, the Hydrogenic and Gaussian profiles yield $|\psi(0)|^2 \approx 0.431$~GeV$^3$ and $0.116$~GeV$^3$, respectively. The present results therefore illustrate that a large difference in this local wave-function quantity does not translate into a comparably large change in the quoted transition form factors. No direct causal relation between the wave-function overlap and the form-factor variation is assumed here. 

\begin{figure*}[t]
    \centering
    \includegraphics[width=0.9\textwidth]{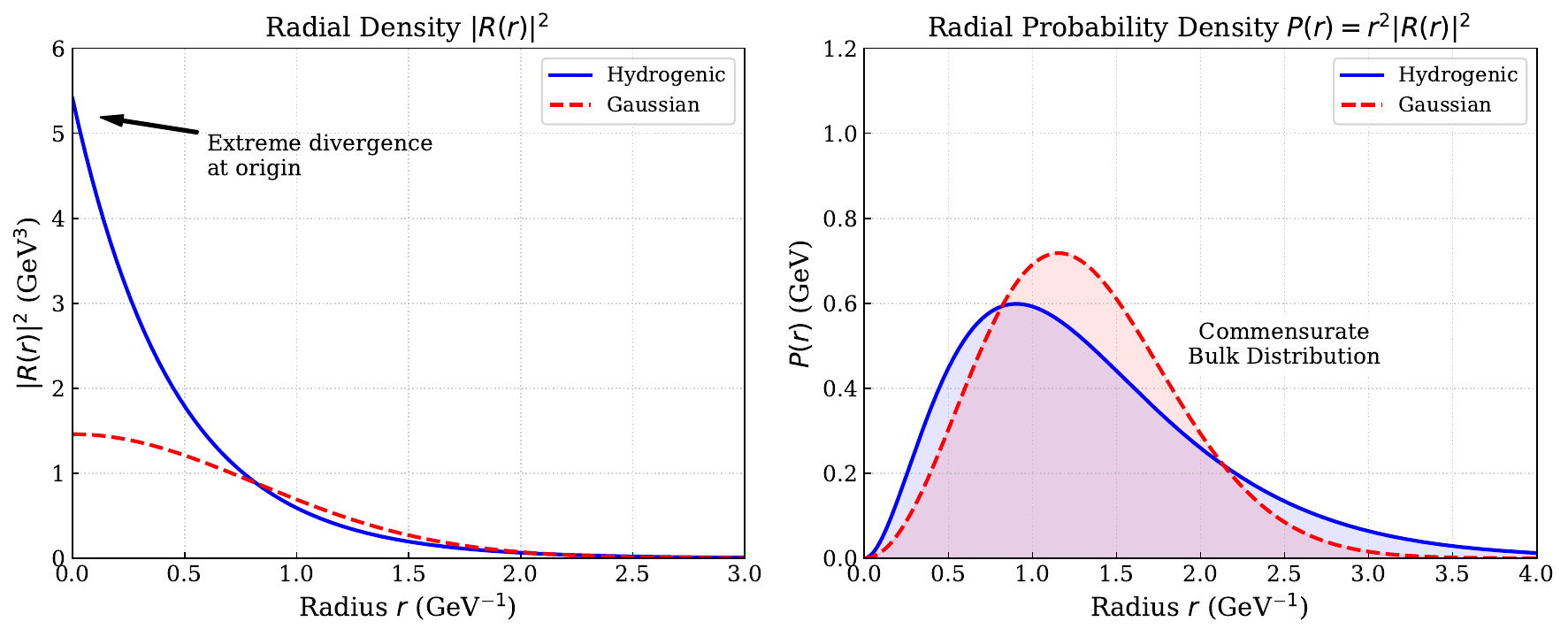}
    \caption{Comparison of the radial structure of the $B_c(1S)$ meson under the Hydrogenic ($\gamma=2.212$) and Gaussian ($\gamma=0.865$) models. The left panel displays the radial density $\vert{}R(r)\vert{}^2$, illustrating the different behavior of the two profiles near the origin. The right panel displays the radial probability density $P(r)=r^2\vert{}R(r)\vert{}^2$, which emphasizes the spatial distribution over radius.}
    \label{fig:probability_density}
\end{figure*}

The similarity of the form-factor values does not persist to the same extent in the integrated physical observables. As Table~\ref{tab:sensitivity} demonstrates, the decay widths and branching ratios exhibit a more pronounced model dependence. For example, the corrected branching ratio for the semileptonic $B_c^- \to J/\psi e^-\bar{\nu}_e$ channel changes from $50.800$\textperthousand\ under the Hydrogenic model to $45.300$\textperthousand\ under the Gaussian model, corresponding to an approximately $11\%$ relative variation. Similarly, the $B_c^- \to \eta_c e^-\bar{\nu}_e$ branching ratio changes from $9.110$\textperthousand\ to $8.270$\textperthousand, corresponding to an approximately $9\%$ relative variation. This difference reflects the dependence of the integrated decay rates on the form factors over the full kinematic range $m_\ell^2 \le q^2 \le (m_{B_c}-m_C)^2$.

The Coulomb-related prescription produces a channel-dependent change in the calculated observables. For example, the corrected $B_c^- \to \eta_c e^-\bar{\nu}_e$ branching ratios lie in the range $8.270$--$9.110$\textperthousand, while the corresponding $J/\psi e^-\bar{\nu}_e$ values lie in the range $45.100$--$50.800$\textperthousand. These results can be compared with the literature values listed in Table~\ref{tab:branching_ratios}; the comparison is presented here as a numerical sensitivity study rather than as a determination of a preferred correction scheme.

The different response of the $\eta_c$ and $J/\psi$ channels to the additional Coulomb prescription also illustrates the sensitivity of the results to how this correction is implemented. A uniform multiplicative factor is applied in the present analysis, whereas the pseudoscalar and vector channels involve different combinations of form factors over their respective kinematic ranges. Moreover, Coulomb-like interactions are already present in the potential-model interaction used to construct the bound-state wave functions. The extent to which the additional multiplicative factor represents an independent contribution therefore remains subject to theoretical interpretation and warrants further investigation.

Within the present implementation, the Coulomb-related prescription can produce sizeable changes in the numerical normalization of the decay observables. Since the correction is motivated by near-threshold dynamics but is applied as a phenomenological multiplicative factor, its use away from the strict near-threshold region should be interpreted cautiously. A momentum-dependent treatment and a more explicit separation of Coulomb effects already contained in the bound-state model from additional threshold contributions would be useful in future studies. The Coulomb-enhanced results are therefore best viewed as illustrating the sensitivity of the observables to the adopted correction prescription rather than as a complete quantitative treatment of near-threshold dynamics.

\section{Conclusion}
\label{sec:conclusion}

This work examined the dependence of $B_c \to \eta_c$ and $B_c \to J/\psi$ decay observables on Hydrogenic and Gaussian radial wave-function models. Numerical QCDSR form-factor information from the literature was used as input and represented through a BCL $z$-parametrization for the decay analysis. The form-factor values at the quoted kinematic point show relatively small model-induced variations, whereas the integrated decay widths and branching ratios exhibit larger variations for several channels. These differences should be interpreted as a model-induced spread rather than as a complete theoretical uncertainty estimate.

The additional Coulomb/Sommerfeld-type prescription considered in the analysis produces further channel-dependent changes in the calculated observables. Because Coulomb-like interactions are already included in the potential-model description, the additional correction is treated as a sensitivity prescription rather than as an independently established contribution. Future studies could examine more momentum-dependent treatments of threshold effects and a more systematic separation of bound-state and Coulomb-enhancement contributions.

\rev{Potential models and QCD sum rules have been widely employed to investigate the properties of conventional and exotic hadronic systems. Their applications extend from ordinary mesons and baryons to multiquark configurations, including tetraquarks, pentaquarks, hexaquarks and dibaryons. For example, Rai et al. investigated various multiquark candidates within a di-hadronic molecular model }\cite{Rai2007}\rev{, while Shah et al. studied the charmed dibaryon systems }($\Sigma_c\Sigma_c$)\rev{ and }($\Sigma_c^*\Sigma_c^*$) \cite{Shah2024}\rev{. In the QCD sum-rule framework, studies have investigated the }$T_{cc}^{+}$\rev{ tetraquark }\cite{Agaev2021}\rev{, hidden-charm pentaquarks }$P_c(4380)$\rev{ and }$P_c(4450)$ \cite{Azizi2017}\rev{, the }$uuddss$\rev{ hexaquark }\cite{Azizi2020}\rev{, and }$\Delta^0\Delta^0$\rev{ dibaryons }\cite{Mutuk2022}\rev{.}

\rev{An important consideration in choosing between these approaches is their computational requirements. QCD sum-rule calculations of transition form factors involve the evaluation of correlation functions, operator product expansions, spectral densities, condensate contributions and Borel transformations, followed by continuum subtraction. Potential-model calculations, on the other hand, employ effective interactions and bound-state wave functions and can provide spectroscopic properties and decay-related quantities through comparatively simpler calculations. Their application to heavy mesons, including the calculation of masses, decay constants, leptonic branching fractions and electromagnetic transitions, has been demonstrated in several studies }\cite{Devlani2012,Devlani2014PoS}\rev{.}

\rev{These considerations motivate the use of the potential-model framework in the present study to investigate transition form factors and the resulting decay observables. While QCD sum rules provide transition form factors in the spacelike region, the potential-model approach offers a simpler way to evaluate the transition matrix elements in the physical kinematic region. The resulting form factors can subsequently be used to determine decay widths and branching fractions, providing a framework that can be extended to other heavy-meson decays and weak transitions involving exotic particles.}

\bibliographystyle{elsarticle-num}
\bibliography{references}

\end{document}